\documentclass[aps,pre,twocolumn,floatfix]{revtex4-1}

\usepackage{graphicx}
\usepackage{amssymb,amsfonts,amsmath}
\usepackage{epsf}
\usepackage{subfigure}
\usepackage{epstopdf}
\usepackage{mathrsfs}

\newcommand{\be}{\begin{equation}}
\newcommand{\ee}{\end{equation}}
\newcommand{\bea}{\begin{eqnarray}}
\newcommand{\eea}{\end{eqnarray}}

\begin{document}

\title{Comment on ``Validity of path thermodynamics in reactive systems''}

\author{Pierre Gaspard}
\affiliation{Center for Nonlinear Phenomena and Complex Systems,\\
Universit\'e Libre de Bruxelles (U.L.B.), Code Postal 231, Campus Plaine, 
B-1050 Brussels, Belgium}

\begin{abstract}
The paper by Malek~Mansour and Garcia [Phys. Rev. E {\bf 101}, 052135 (2020)]
is shown to be based on misconceptions in the stochastic formulation of chemical thermodynamics in reactive systems.  Their erroneous claims, asserting that entropy production cannot be correctly evaluated using path probabilities whenever the reactive system involves more than one elementary reaction leading to the same composition changes, are refuted.
\end{abstract}


\maketitle

\section{Refutation}
\label{Refutation}

In Ref.~\cite{MG20}, Malek Mansour {\it et al.} have raised doubts on the validity of the stochastic approach, they call ``path thermodynamics'', to evaluate thermodynamic quantities (especially, entropy production) using path probabilities in reactive systems.  In particular, they claim that ``nowadays the general belief is that path thermodynamics is the ultimate theoretical formalism for physicochemical systems ranging from macroscopic to nanometer scale. [...] Yet, we shall prove that the resulting properties will be wrong whenever the reactive system involves more than one elementary reaction leading to the same composition changes''.  Similar criticisms have already been expressed by Malek Mansour {\it et al.} in Ref.~\cite{MB17}.

The aim of this Comment is to show that such criticisms are ill founded because they result from basic misconceptions in the stochastic formulation of chemical thermodynamics in reactive systems.

The reactive systems that are here considered are networks of elementary chemical reactions $\rho=1,2,\dots,r$:
\be
\sum_{i=1}^a \nu_{\rho i}^{(-)}\; {\rm A}_i + \sum_{j=1}^c \nu_{\rho j}^{(-)}\; {\rm X}_j 
\ \underset{W_{-\rho}}{\overset{W_{+\rho}}{\rightleftharpoons}} \ 
\sum_{i=1}^a \nu_{\rho i}^{(+)}\; {\rm A}_i + \sum_{j=1}^c \nu_{\rho j}^{(+)}\; {\rm X}_j \, ,
\label{reactions}
\ee
where $\{ {\rm A}_i \}_{i=1}^a$ are reactant and product molecular species, $\{ {\rm X}_j \}_{j=1}^c$ intermediate molecular species, and
\be
\nu_{\rho j}\equiv \nu_{\rho j}^{(+)}-\nu_{\rho j}^{(-)} 
\label{stoichio}
\ee
are the stoichiometric coefficients of species $j$ in the reaction $\rho$.  The rates of the forward and backward reactions $\pm\rho$ are respectively denoted $W_{\pm\rho}$.  For a system of volume~$V$, the molecular concentrations are given by $a_i=A_i/V$ and $x_j=X_j/V$, where $A_i$ and $X_j$ are the numbers of molecules of corresponding species in the reactive system.  This latter is supposed to be homogeneous because of efficient mixing.

If the intermediate species  are assumed to have lower concentrations than reactants and products ($a_i\gg x_j$) either at initial time or by supply from outside, the system may reach a stationary state determined by the reactant and product concentrations.  This stationary state is either the equilibrium steady state if the conditions of detailed balance are satisfied, according to which every pair of forward and backward elementary reactions $\pm\rho$ are in balance; or a nonequilibrium steady state (NESS) if this is not the case.

At the mesoscopic level of description, the numbers $A_i$ and $X_j$ of molecules are erratically changing in time because of reactive events occurring at random upon inelastic collisions between molecules.  Such physicochemical systems can be described in the framework of the theory of stochastic processes.  If the reactive events are fast enough with respect to the mean time between their occurrences, these stochastic processes may be taken as continuous-time discrete-state Markov processes defined in terms of transition rates $W_{\pm\rho}$.  The theory of such stochastic Markovian processes has been much developed since classic works by Kolmogorov~\cite{K31} and Feller~\cite{F49,F57}.  A famous theorem by them states that the probabilities of such a Markov process are given by the solutions of a master equation and that these solutions are essentially unique.  This master equation can be written down in terms of the transition rates $W_{\pm\rho}$ and it can be solved to obtain the probability distributions of the random variables at given instant of time, as well as the probabilities of the paths taken by these variables at successive instants of time.

In Ref.~\cite{MG20}, Malek Mansour {\it et al.} claim that ``... we are entirely free to define a ``path" any way we want.  But then it is {\it not} always possible to associate a stochastic process to an arbitrary constructed path.  This misinterpretation of the Kolmogorov's theorem is at the origin of the wrong result of Gaspard and Andrieux'' \cite{G04,AG0408}.  The fallacy of this statement is that Malek Mansour {\it et al.}~\cite{MG20} here suppose that the uniqueness theorem of Kolmogorov (and Feller) imposes some limitations in the description of a physicochemical system by different possible stochastic processes.  Actually, the stochastic processes concerned by the uniqueness theorem of Kolmogorov (and Feller) are defined by choosing given random variables, although different sets of random variables may be chosen for the description of a physicochemical system, thus leading to the definition of different possible stochastic processes.  Therefore, the uniqueness theorem does not establish the uniqueness of the stochastic process associated with a given physicochemical system.  The misuse of the uniqueness theorem in Ref.~\cite{MG20} is misleading Malek~Mansour {\it et al.} to incorrect conclusions that will now be refuted.

As a matter of fact, different sets of random variables may be chosen to describe different properties of interest in reactive systems (as well as in stochastic mechanical systems).  These properties may be the numbers~$X_j$ of molecules of the intermediate species in the reaction network~(\ref{reactions}), or the amounts of reactants and chemical free energy that are consumed when the reaction network is driven out of equilibrium, or also the entropy that is produced due to dissipation.  Establishing the balances of energy and entropy is the goal of thermodynamics.  Besides, depending on the set of random variables chosen to describe physicochemical systems, the stochastic process may be reversible or not, whenever the system is at equilibrium or not, as can be illustrated in many physicochemical systems.  The choice of random variables is thus crucial for the description of nonequilibrium systems and their thermodynamics.  Fully aware of these fundamental issues, the author of this Comment proposed in 2004 to describe the time evolution of a stochastic reactive system in terms of the paths
\be
{\cal X}(t) = \pmb{X}_0\; {\overset{\rho_1}\longrightarrow} \; \pmb{X}_1
\; {\overset{\rho_2}\longrightarrow} \; \pmb{X}_2\;  {\overset{\rho_3}\longrightarrow} \; \cdots\;
{\overset{\rho_n}\longrightarrow}\; \pmb{X}_n \; ,
\label{path}
\ee
where $\pmb{X}_l$ are the numbers of molecules of intermediate species at the successive times $t_1<t_2<\cdots<t_n$ and $\rho_l$ are the reactions occurring along the path, while the initial condition is sampled according to the stationary probability distribution $P_{\rm st}(\pmb{X}_0)$ \cite{G04}.  The successive random reactive events can be generated with Gillespie's algorithm~\cite{Gillespie76,Gillespie77} based on the following rates of the elementary reactions in the network~(\ref{reactions}):
\be\label{rates}
W_{\rho}(\pmb{X}\vert\pmb{X'}) \qquad\mbox{for the transition} \quad \pmb{X} {\overset{\rho}\longrightarrow}\, \pmb{X'}=\pmb{X}+ \pmb{\nu}_{\rho}
\ee
with $\rho=\pm 1,\pm 2,\dots,\pm r$ and the stoichiometric coefficients $\pmb{\nu}_{\rho}=\{\nu_{\rho j}\}_{j=1}^{c}$.  Far from being superfluous, the reaction sequence $\{\rho_1,\rho_2,\dots,\rho_l,\dots,\rho_n\}$ is playing a key role and it must be specified to be consistent with the Gillespie algorithm simulating the reactive events.  Indeed, Gillespie's algorithm provides an exact Monte Carlo method for simulating the stochastic time evolution of coupled chemical reactions described by jump Markov processes \cite{Gillespie76,Gillespie77}.  Moreover, it is worth pointing out that the aim of the original algorithm is the simulation of the reactive events, instead of the jumps of the molecular numbers~$\pmb{X}$.  

Apparently, Malek Mansour {\it et al.} in Refs.~\cite{MG20,MB17} have overlooked this essential aspect in the definition of the paths and they have wrongly assumed that the paths are defined in Ref.~\cite{G04} as
\be
{\cal X}_0(t) = \pmb{X}_0 \longrightarrow \pmb{X}_1
\longrightarrow \pmb{X}_2 \longrightarrow \cdots
\longrightarrow \pmb{X}_n
\label{path0}
\ee
by omitting the reaction sequence.  The stochastic process resulting from this omission is defined in terms of the sole random variables $\pmb{X}_l$ that are the molecular numbers of the intermediate species.  The rates of this other stochastic process are given by summing the rates of all the elementary reactions having equal stoichiometric coefficients for the intermediate species considered, according to
\be\label{rates0}
W(\pmb{X}\vert\pmb{X'}) \equiv \sum_{\rho} \delta_{\pmb{X'}-\pmb{X},\pmb{\nu}_{\rho}} W_{\rho}(\pmb{X}\vert\pmb{X'})  \, .
\ee
This other stochastic process is thus different from the previous one if the reactive system involves more than one elementary reaction leading to the same composition changes.  For numerical purposes, the rates~(\ref{rates0}) define a different Gillespie algorithm, which can have the advantage of speeding up simulations.  However, the limitation is that this other stochastic process does not contain enough information to evaluate the entropy production, which instead requires the knowledge of the elementary chemical reactions taking place in the reactive system \cite{JVN84}.  The theorem by Kolmogorov and Feller certainly guarantees the uniqueness of the path probabilities $P_{\rm st}[{\cal X}(t)]$ and $P_{\rm st}[{\cal X}_0(t)]$ associated with each one of the two stochastic processes, but, contrary to the erroneous claims by Malek~Mansour {\it et al.} in Ref.~\cite{MG20}, this theorem does not disprove the existence of either one or the other of those processes.

Now, we continue with the issue of evaluating the entropy production using path probabilities.
Inspired by Ref.~\cite{LS99}, the author of the present Comment proposed in Ref.~\cite{G04} that entropy production can be evaluated in reactive systems using the ratio of the probability $P_{\rm st}[{\cal X}(t)]$ of the path~(\ref{path}) to the probability of its time reversal
\be
{\cal X}^{\rm R}(t) =\pmb{X}_n\; {\overset{-\rho_n}\longrightarrow} \;
\cdots \; {\overset{-\rho_3}\longrightarrow}\; \pmb{X}_2 \;
{\overset{-\rho_2}\longrightarrow} \; \pmb{X}_1\; {\overset{-\rho_1}\longrightarrow} \; \pmb{X}_0 \; .
\label{R.path}
\ee
The logarithm of this ratio can be written as
\be
\ln \frac{P_{\rm st}[{\cal X}(t)]}{P_{\rm st}[{\cal X}^{\rm R}(t)]} = Z(t) + \ln \frac{P_{\rm st}(\pmb{X}_0)}{P_{\rm st}(\pmb{X}_n)}
\label{Sigma}
\ee
in terms of the quantity
\be
Z(t) \equiv \ln\prod_{l=1}^n \frac{W_{\rho_l}(\pmb{X}_{l-1}\vert\pmb{X}_l)}{W_{-\rho_l}(\pmb{X}_l\vert\pmb{X}_{l-1})} \, .
\label{Z}
\ee
In the long-time limit, the last term in Eq.~(\ref{Sigma}) becomes negligible in front of $Z(t)$.  This latter is linearly growing with time under nonequilibrium conditions, so that the mean rate of entropy production is given by
\be\label{EPR-Z}
\frac{d_{\rm i}S}{dt}\bigg\vert_{\rm st} =R \equiv \lim_{t\to\infty} \frac{1}{t} \, \langle Z(t)\rangle
\ee
in terms of the mean value of the quantity~(\ref{Z}) with respect to the path probability of the stochastic process,
\be
\langle Z(t)\rangle \equiv \sum_{{\cal X}(t)} P_{\rm st}[{\cal X}(t)] \; Z(t) \, ,
\label{aver}
\ee
Boltzmann's constant being set equal to one.
As proved in Ref.~\cite{G04}, this method leads to the standard expression of the entropy production rate in reactive systems \cite{JVN84}
\bea\label{EPR-std}
\frac{d_{\rm i}S}{dt}\bigg\vert_{\rm st} &=& \sum_{\pmb{X}}\sum_{\rho=1}^{r}  \Big[ W_{\rho}(\pmb{X}-\pmb{\nu}_{\rho}\vert\pmb{X})\; P_{\rm st}(\pmb{X}-\pmb{\nu}_{\rho})\nonumber\\
&&\qquad\qquad - W_{-\rho}(\pmb{X}\vert\pmb{X}-\pmb{\nu}_{\rho})\; P_{\rm st}(\pmb{X})\Big]  \nonumber\\
&& \times \ln \frac{W_{\rho}(\pmb{X}-\pmb{\nu}_{\rho}\vert\pmb{X})\; P_{\rm st}(\pmb{X}-\pmb{\nu}_{\rho})}{W_{-\rho}(\pmb{X}\vert\pmb{X}-\pmb{\nu}_{\rho})\; P_{\rm st}(\pmb{X})} \ge 0 \, . \qquad
\eea

In contradistinction with what is written in Ref.~\cite{G04}, Malek Mansour {\it et al.} have supposed in  Refs.~\cite{MG20,MB17} that the paths~(\ref{path0}) omitting the reaction sequence and their time reversal would have been considered in Ref.~\cite{G04}.  This allegation is misleading Malek~Mansour {\it et al.} to consider the different quantity
\be
\ln \frac{P_{\rm st}[{\cal X}_0(t)]}{P_{\rm st}[{\cal X}^{\rm R}_0(t)]} = Z_0(t) + \ln \frac{P_{\rm st}(\pmb{X}_0)}{P_{\rm st}(\pmb{X}_n)}
\label{Sigma0}
\ee
with
\be
Z_0(t) \equiv \ln\prod_{l=1}^n \frac{W(\pmb{X}_{l-1}\vert\pmb{X}_l)}{W(\pmb{X}_l\vert\pmb{X}_{l-1})} \, ,
\label{Z0}
\ee
which is defined in terms of the rates~(\ref{rates0}) of the stochastic process~(\ref{path0}), instead of the rates~(\ref{rates}) of the stochastic process~(\ref{path}).  Given that the assumptions are different, it is not a surprise that this is also the case for the conclusions that are reached.

Actually, the mean growth rate of the quantity~(\ref{Z0}) is always smaller than or equal to the entropy production rate given by Eqs.~(\ref{EPR-Z}) and~(\ref{EPR-std}):
\be\label{bound-EPR}
R_0 \equiv \lim_{t\to\infty} \frac{1}{t} \, \langle Z_0(t)\rangle \le R = \frac{d_{\rm i}S}{dt}\bigg\vert_{\rm st} \, .
\ee
If the stochastic process of the paths~(\ref{path0}) is reversible although the system is out of equilibrium, the rate $R_0$ in Eq.~(\ref{bound-EPR}) is equal to zero, confirming that the paths~(\ref{path0}) may not be used to correctly evaluate the entropy production rate.

As proved in Ref.~\cite{G04}, it is the rate $R$ in Eq.~(\ref{EPR-Z}) that gives the correct value of the entropy production rate in reactive systems since we need to consider the paths~(\ref{path}) in order to identify the transitions associated with the elementary chemical reactions, as required by the principles of chemical thermodynamics \cite{JVN84}.

Finally, an attempt is made by Malek Mansour {\it et al.} in Eqs.~(14)-(16) of Ref.~\cite{MG20} trying to disprove the fluctuation theorem given by their Eq.~(7) for the quantity~$Z(t)$. What is wrong with this attempt is that it is again confusing the properties of the two different stochastic processes associated with the paths defined by either~(\ref{path}) or~(\ref{path0}).  On the one hand, the quantity $Z(t)$ in Eq.~(14) of Ref.~\cite{MG20} can only be defined for the paths~${\cal X}(t)$ defined here above by Eq.~(\ref{path}) including the sequence $\{\rho_l\}_{l=1}^n$ of elementary reactions in the path, but the probability of this path is not equal to the probability of its time reversal, $P_{\rm st}[{\cal X}(t)] \ne P_{\rm st}[{\cal X}^{\rm R}(t)]$, in contradiction with the so-called ``fundamental relation'' used to obtain Eq.~(16) of Ref.~\cite{MG20}.  On the other hand, for the Schl\"ogl model \cite{Schl71,Schl72} considered in this part of Ref.~\cite{MG20}, the stochastic process based on the paths~${\cal X}_0(t)$ defined here above by Eq.~(\ref{path0}) omitting the reaction sequence is a reversible process, for which $P_{\rm st}[{\cal X}_0(t)] = P_{\rm st}[{\cal X}^{\rm R}_0(t)]$ indeed holds.  However, for the stochastic process corresponding to the paths~${\cal X}_0(t)$ where the reaction sequence is omitted, the quantity $Z(t)$ defined in Eq.~(14) of Ref.~\cite{MG20} is undefined since the sequence $\{\rho_l\}_{l=1}^n$ of elementary reactions is not specified, contradicting the premises.  Consequently, in both alternatives, Eq.~(16) of Ref.~\cite{MG20} does not invalidate the fluctuation theorem given by Eq.~(7) of the same paper.

Thus, the claims by Malek Mansour {\it et al.} in Ref.~\cite{MG20}, according to which the results of Refs.~\cite{G04,AG0408} would be wrong, are inconsistent due to basic mistakes in the stochastic formulation required to evaluate entropy production in reactive systems.


\section{Illustrative example}
\label{Ex}

These issues are easily illustrated with the following simple reaction network,
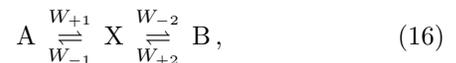
\be
{\rm A} \, \underset{W_{-1}}{\overset{W_{+1}}{\rightleftharpoons}}\, {\rm X} \, \underset{W_{+2}}{\overset{W_{-2}}{\rightleftharpoons}}\, {\rm B} \, , \label{react-Ex}
\ee
where A and B are the reactant or product species, while X is the intermediate species.
The numbers of these species satisfy the conservation law $A(t)+B(t)+X(t)=A(0)+B(0)+X(0)$ at every time $t$.  The transition rates of this reaction network are given by
\bea
&& W_{+1} = k_{+1} \; A \, , \quad W_{-1} = k_{-1} \; X \, , \nonumber\\
&& W_{+2} = k_{+2} \; B \, , \quad W_{-2} = k_{-2} \; X\, . \label{rates-Ex}
\eea

The Gillespie algorithm simulating this process is based on the distinct rates $W_{\pm 1}$ and $W_{\pm 2}$ and it generates the paths 
\begin{equation}
{\cal X}(t) = X_0\; {\overset{\rho_1}\longrightarrow}\;X_1\;
{\overset{\rho_2}\longrightarrow} \; X_2 \; {\overset{\rho_3}\longrightarrow}\; \cdots\;
{\overset{\rho_n}\longrightarrow}\; X_n
\label{path-Ex}
\end{equation}
specified not only by the sequence of molecule numbers $X_l$, but also the reactive events $\rho_l\in\{\pm1,\pm2\}$.  The knowledge of the reaction sequence allows us to reconstruct the time evolution of the number $A$ of reactant or product molecules that is consumed during the process.  The number of molecules~B is given by $B(t)=A(0)+B(0)+X(0)-A(t)-X(t)$.  Therefore, every path~(\ref{path-Ex}) is equivalent to the path
\bea
{\cal X}(t) &=& (A_0,X_0) \longrightarrow (A_1,X_1)
\longrightarrow (A_2,X_2) \longrightarrow \nonumber\\
&&\cdots \longrightarrow (A_n,X_n) \, .
\label{path-Ex-A-X}
\eea
The stochastic process defined for the paths~(\ref{path-Ex}) is thus identical to the stochastic process defined for the paths~(\ref{path-Ex-A-X}) in terms of the random variables $A$ and $X$.  The master equation of this stochastic process is given by
\bea
&&\frac{d}{dt}P(A,X,t) =  W_{+1}(A+1,X-1) \, P(A+1,X-1,t) \nonumber\\
&&\qquad\qquad + W_{-1}(A-1,X+1) \, P(A-1,X+1,t) \nonumber\\
&&\qquad\qquad + W_{+2}(A,X-1) \, P(A,X-1,t) \nonumber\\
&&\qquad\qquad + W_{-2}(A,X+1) \, P(A,X+1,t) \nonumber\\
&&\qquad\qquad - \Big[ W_{+1}(A,X) + W_{-1}(A,X) \nonumber\\
&&\qquad\qquad + W_{+2}(A,X)  + W_{-2}(A,X) \Big] P(A,X,t) \, , 
\label{eq-A-X}
\eea
ruling the time evolution of the probability $P(A,X,t)$ that the system contains the numbers $A$ and $X$ of molecules at time $t$.  

Now, we may consider the stochastic process associated with the paths
\begin{equation}
{\cal X}_0(t) \; = \; X_0\; {\longrightarrow}\; X_1
\; {\longrightarrow}\; X_2 \; {\longrightarrow}\; \cdots\;
{\longrightarrow}\; X_n
\label{path-Ex-X}
\end{equation}
obtained from the paths~(\ref{path-Ex}) by omitting the reaction sequence $\{\rho_1,\rho_2,\dots,\rho_n\}$ or, equivalently, from the paths~(\ref{path-Ex-A-X}) by erasing the random variables $\{A_0,A_1,\dots,A_n\}$. Clearly, the information contained in the paths~(\ref{path-Ex-X}) is too limited to reconstruct the paths~(\ref{path-Ex}) or~(\ref{path-Ex-A-X}) and thus to determine the sequence of elementary chemical reactions that have been followed during the process, in order to evaluate the entropy production.  

Here, it is often assumed that the reactant and product species are much more abundant than the intermediate species, so that their concentrations $a=A/V$ and $b=B/V$ remain essentially constant during the process.  Under such circumstances, the rates $W_{+1}\simeq k_{+1} a V$ and $W_{+2}\simeq k_{+2} b V$ no longer depend on the random variable $A$ in the master equation~(\ref{eq-A-X}).  We may thus deduce the following master equation ruling the time evolution of the marginal probability distribution $P(X,t) \equiv \sum_{A} P(A,X,t)$ for the sole random variable $X$,
\bea
&& \frac{d}{dt}P(X,t) =  W_{+}(X-1)\, P(X-1,t) \nonumber\\
&&\qquad\qquad\quad + W_{-}(X+1)\, P(X+1,t)\nonumber\\
&&\qquad\qquad\quad -\left[ W_{+}(X)+W_{-}(X) \right] P(X,t) \qquad
\label{eq-X}
\eea
with the cumulative transition rates $W_{\pm} \equiv W_{\pm 1}+W_{\pm 2}$.  The paths of this reduced stochastic process are given by Eq.~(\ref{path-Ex-X}).

Now, we can compare the rates $R$ and $R_0$ respectively given in terms of the quantities~(\ref{Z}) and~(\ref{Z0}).  Under the conditions where $A\simeq aV$ and $B\simeq bV$ are constant, the stationary solution of Eq.~(\ref{eq-X}) is a Poisson distribution of mean value $\langle X\rangle_{\rm st}=V x_{\rm st}$ corresponding to the stationary concentration $x_{\rm st}=(k_{+1}a+k_{+2}b)/(k_{-1}+k_{-2})$ of the intermediate species.  In this stationary state, the macroscopic rates $w_{\rho}\equiv W_{\rho}/V$ are satisfying the condition $w_{+1}+w_{-1}=w_{+2}+w_{-2}$.  In the limit where the volume $V$ is large enough, the Poisson distribution becomes Gaussian and peaked around $\langle X\rangle_{\rm st}=V x_{\rm st}$, so that the mean growth rate of the quantity~(\ref{Z}) is obtained from Eq.~(\ref{EPR-std}), giving
\bea
R = \frac{d_{\rm i}S}{dt}\bigg\vert_{\rm st} &\simeq& V \bigg[ (w_{+1}-w_{-1}) \ln \frac{w_{+1}}{w_{-1}} \nonumber\\
&&\ \ + (w_{+2}-w_{-2}) \ln \frac{w_{+2}}{w_{-2}} \bigg]_{\rm st} \, .
\label{EPR-Ex}
\eea
This entropy production rate is equal to zero at equilibrium when the conditions of detailed balance hold: $w_{+1}=w_{-1}$ and $w_{+2}=w_{-2}$.  Otherwise, it takes a positive value in NESS.  The stochastic process associated with the paths~(\ref{path-Ex}) is thus reversible at equilibrium, but not reversible in NESS.

However, the mean growth rate~(\ref{bound-EPR}) of the quantity~(\ref{Z0}) tends towards the value
\be
R_0 \simeq V \, (w_{+}-w_{-}) \ln \frac{w_{+}}{w_{-}} 
\label{R0-Ex-X}
\ee
with the cumulative reaction rates $w_{\pm} \equiv w_{\pm 1} + w_{\pm 2}$.  Under stationary conditions where $w_+=w_-$, the mean growth rate~(\ref{R0-Ex-X}) is thus vanishing in consistency with the reversibility of the stochastic process associated with the paths~(\ref{path-Ex-X}) for the reaction network~(\ref{react-Ex}), although the system may be in NESS.

These considerations demonstrate that several stochastic processes may describe a reactive system and that the entropy production is correctly evaluated using one giving complete information on the reaction sequence.


\section{Conclusion}
\label{conclusion}

As explained in this Comment, the doubts and criticisms expressed by Malek~Mansour {\it et al.} in Refs.~\cite{MG20,MB17} about the validity of the stochastic approach to evaluate entropy production in reactive systems are ill founded. 

Contrary to what is claimed in Refs.~\cite{MG20,MB17}, several Markov processes can be defined for a given reactive system, depending on the set of random variables that are chosen.  Each of these stochastic processes is associated with a specific Gillespie algorithm and, in parallel, the theorem of Kolmogorov and Feller establishes the uniqueness of the associated path probabilities.  

Since the entropy production of reactive systems is defined in terms of the elementary chemical reactions~\cite{JVN84}, it is the stochastic process associated with the Gillespie algorithm simulating the successive random events of these elementary chemical reactions that should be considered to evaluate entropy production in the stochastic approach.

Omitting reaction sequences leads to incorrect values for entropy production.  This is expected since this omission is further coarse graining the description of the process, up to being so coarse that its nonequilibrium character may become unobservable.  This general feature can also be illustrated in other physicochemical systems \cite{BEG11}. For instance, Brownian motion in an external force can be described by the stochastic process defined with the random variables of position and velocity, which is ruled by Kramers' master equation \cite{K40}.   However, the stochastic process for the sole random variable given by the velocity is in one-to-one correspondence with the Ornstein-Uhlenbeck stochastic process, which is known to be reversible.  The situation is thus analogous to what happens for the reaction network of Sec.~\ref{Ex}.  In both systems, the observation of the sole random variable defining a reversible process does not give direct evidence for the nonequilibrium character of the system.


\section*{Acknowledgments}

This research is financially supported by the Universit\'e Libre de Bruxelles (ULB) and the Fonds de la Recherche Scientifique~-~FNRS under the Grant PDR~T.0094.16 for the project ``SYMSTATPHYS".


\end{document}